\documentclass[]{sig-alternate}

\pdfoutput=1
\usepackage{tablefootnote}
\usepackage{graphicx, subfigure, multirow, times, balance, color, hyperref}
\usepackage{balance, url, amsfonts, verbatim, mathtools, paralist}
\usepackage{graphicx, xspace, subfigure,grffile, array}
\usepackage{comment}
\usepackage{algorithm, algpseudocode}
\usepackage[letterpaper,hmargin={1in,1in},vmargin={1in,1in}]{geometry}
\usepackage[small]{caption}
\usepackage{listings}
\makeatletter
\newcommand\subparagraph{%
  \@startsection{subparagraph}{5}
  {\parindent}
  {3.25ex \@plus 1ex \@minus .2ex}
  {-1em}
  {\normalfont\normalsize\bfseries}}
\makeatother
\usepackage[small,compact]{titlesec}
\let\subparagraph\relax
\usepackage[sort]{cite}
\usepackage{mdframed}

\hypersetup{
    colorlinks=true,
    linkcolor=black,
    citecolor=black,
    filecolor=black,
    urlcolor=black,
}

\usepackage{mdwlist}

\newcommand{\bi}{\begin{itemize}}
\newcommand{\ei}{\end{itemize}}

\newcommand{\eg}{{\it e.g.,}\xspace}
\newcommand{\ie}{{\it i.e.,}\xspace}
\newcommand{\lxb}{LXB\xspace}
\newcommand{\lxbs}{LXBs\xspace}

\newcommand{\anlxb}{an LXB\xspace}
\algnewcommand{\LineComment}[1]{\State \(\triangleright\) #1}

\newcommand{\allnotes}[1]{}
\renewcommand{\allnotes}[1]{\textit{#1}} 

\newcommand{\eat}[1]{}

\def\squarebox#1{\hbox to #1{\hfill\vbox to #1{\vfill}}}

\begin{document}

\title{Recursive SDN for Carrier Networks}

\author{
James McCauley$^{\ddag\blacktriangle}$, Zhi Liu$^\dag$, Aurojit Panda$^\ddag$\\
Teemu Koponen$^\diamond$, Barath Raghavan$^{\blacktriangle}$, Jennifer Rexford$^\blacklozenge$, Scott Shenker$^{\ddag\blacktriangle}$\\
\\
$^\ddag$UC Berkeley, $^\blacktriangle$ICSI, $^\dag$Tsinghua University, $^\diamond$Styra, $^\blacklozenge$Princeton
}

\date{}
\maketitle
\begin{abstract}
{\em Control planes for global carrier networks should be programmable (so that new functionality can be easily introduced) and scalable (so they can handle the numerical scale and geographic scope of these networks). Neither traditional control planes nor new SDN-based control planes meet both of these goals. In this paper, we propose a framework for recursive routing computations that combines the best of SDN (programmability) and traditional networks (scalability through hierarchy) to achieve these two desired properties. Through simulation on graphs of up to 10,000 nodes, we evaluate our design's ability to support a variety of routing and traffic engineering solutions, while incorporating a fast failure recovery mechanism.}
\end{abstract}



\section{Introduction}
\label{sec:intro}

The goal of Software-Defined Networking (SDN) is to make network control planes programmable. While SDN has made great progress in various contexts, most notably within datacenters and in private WANs that interconnect datacenters, there has been surprisingly little published work on using SDN in a more traditional networking context: that of global-scale carrier networks (such as operated by Deutsche Telekom, France Telecom, NTT, AT\&T, and others). These carrier networks are far more geographically dispersed than datacenter networks (by roughly four orders of magnitude)\footnote{Datacenter diameters are significantly less than a mile while that of global carrier networks can reach ten thousand miles.}, while having far more nodes than the global networks that are used solely to interconnect those datacenters (by roughly three orders of magnitude).\footnote{Inter-datacenter networks typically have fewer than 50 nodes, while carrier networks can have on the order of 50,000 nodes.} There are SDN designs that can handle large numbers of nodes (\eg Kandoo \cite{kandoo}), and SDN designs that can handle global networks that interconnect a limited number of datacenters (\eg B4 \cite{b4}), but to our knowledge there are no SDN designs that simultaneously handle both the numerical scale and geographic scope of today's carrier networks.

The challenge -- heretofore unmet in the global carrier context -- is to provide global programmability while retaining {\em locality-of-control}, by which we mean providing rapid responses to events (when they can be handled locally) and preserving fate-sharing (not unnecessarily relying on distant parts of the network). Typical SDN designs involve a logically centralized (but often replicated) controller. Applying this approach to carrier networks would violate both requirements of locality-of-control: (i) the control plane would incur significant round-trip (or controller consensus) delays for all control decisions, and (ii) the control plane would rely on connectivity to this logical controller.

Thus, for global carrier networks we must extend the SDN paradigm. To that end, we propose a recursive approach to SDN routing\footnote{We use the term ``routing'' broadly to encompass the choice of routes in unicast, multicast/anycast, and traffic engineering.} -- called Recursive SDN (RSDN) -- that leverages the hierarchical structure of carrier networks to achieve the programmability of SDN networks while retaining the scalability and locality-of-control (through hierarchy) of legacy networks. In RSDN, each level of the route computation acts on a set of aggregates (called logical cross-bars, or \lxbs), and then communicates a summary of the results to the appropriate parent and child \lxbs.  This approach provides programmability while (i) limiting the number of nodes any individual route computation has to handle (thereby providing scalability) and (ii) making route computations as local as possible, only involving the affected \lxb and, recursively, its children (thereby preserving locality-of-control).

RSDN is not just another hierarchical routing algorithm. Instead, it is a recursive {\em programming framework}; the RSDN programmatic API enables one to build a wide variety of recursive designs for unicast routing, multicast/anycast routing, and traffic engineering.

However, RSDN does more than merely facilitate route {\em computation}. Because availability is such a crucial requirement for carrier networks, RSDN  incorporates a mechanism for rapid and local recovery from failures that is independent of the particular routing algorithm. Typical routing designs (notable exceptions being F10 \cite{f10} and DDC \cite{ddc}) have a fast failover option that handles only a narrow range of failure cases (such as single link failures), and then resort to full route recomputation (using the normal route computation engine) to handle more general failure cases. RSDN starts with standard link protection but then extends this to a more general {\em network repair} mechanism that can handle all failure cases (by having the network with failed links support a virtual version of the failure-free network).  This network repair process is not as immediate as local failover, but it is far faster than global route computations because it need only consider the local region in which the failure occurs. As we demonstrate later, as compared to merely using link protection, network repair decreases routing failures by several orders of magnitude.\footnote{A routing failure is where two endpoints are physically connected but routing is not providing a usable path due to failure.}
 The novelty here is that RSDN's repair mechanism is (i) able to handle all failure scenarios, which goes far beyond techniques such as FRR \cite{rfc4090} and (ii) built into the RSDN framework so that individual routing designs built within RSDN need not  incorporate fast failover techniques themselves. Moreover, by having a built-in rapid repair mechanism, RSDN removes the requirement that global route computation be fast, or be optimized for incremental computations, thereby allowing for a broader class of route computations and easier implementation of them.

It is important to clarify that RSDN focuses only on edge-to-edge packet delivery, which is only a small subset of control plane functionality. In addition to routing, network control planes are often used to enforce policies (\eg through the use of ACLs), create virtual networks (for various tenants), and invoke middlebox functionality (by ensuring packets traverse the appropriate middleboxes). {\em RSDN does none of these additional tasks.} We follow the approach espoused in \cite{fabric} and \cite{shenkerstanfordtalk} in which all non-routing functionality is implemented at the network edge through a variety of known techniques (and thus need not be controlled by RSDN). We adopt this approach because it can support the necessary functionality while creating a network modularity with a clean separation of concerns.

In the next two sections, we give a high-level overview of our design and then some necessary context. The following three sections present detailed design and performance evaluations of two unicast routing algorithms (Section \ref{sec:unicast}), two traffic engineering mechanisms (Section \ref{sec:te}) and the built-in network repair mechanism (Section \ref{sec:repair}). We end with a discussion of miscellaneous issues (Section \ref{sec:misc}) and a few concluding comments (Section \ref{sec:conclusion}).


\section{Design Overview}
\label{sec:spaces}

{\bf Recursive structure:}
In designing RSDN, we exploit the locality that is found in almost all networks (and particularly in carrier networks), with links more likely to exist between nearby network nodes than remote ones. We leverage this locality by clustering the network into aggregates, which we call logical cross-bars (\lxbs); these \lxbs act like switches, in that they have a set of external ports, and can provide transit between those external ports (and where every port attaching a host is thus considered an external port). We repeat this process of aggregation on the \lxbs to build a hierarchical structure with each k-level \lxb being comprised of multiple (k+1)-level \lxbs,
and with links between the \lxbs on the same level (or tier). For example, the levels of aggregation may include: PoP and/or datacenter, access network, regional network, and country (or continental) network. This kind of aggregation is standard in networking (and, in particular, is used in approaches such as PNNI \cite{pnni} to aggregate the topology), but here we are leveraging this hierarchical structure to form an explicitly recursive SDN control plane.

{\bf Controllers:} We associate  a  logical  controller  with  each LXB.  Within these logical controllers,  developers provide recursive control programs -- which we term \textit{control logics} -- to perform control plane tasks (\eg implementing a routing algorithm).  Each controller is logically a single entity, but should typically be replicated across multiple physical machines for reliability.\footnote{Notably, our prototype implementation does not implement this, as it is orthogonal to RSDN and is a largely solved problem, \eg by ONOS\cite{onos} and standard availability techniques such as hot/warm/cold standbys.} When we say that an LXB computes something or sends a message, we mean that control logics within a logical controller take these actions on its behalf.

\begin{figure}[t]
\centering
\includegraphics[width=3.09in]{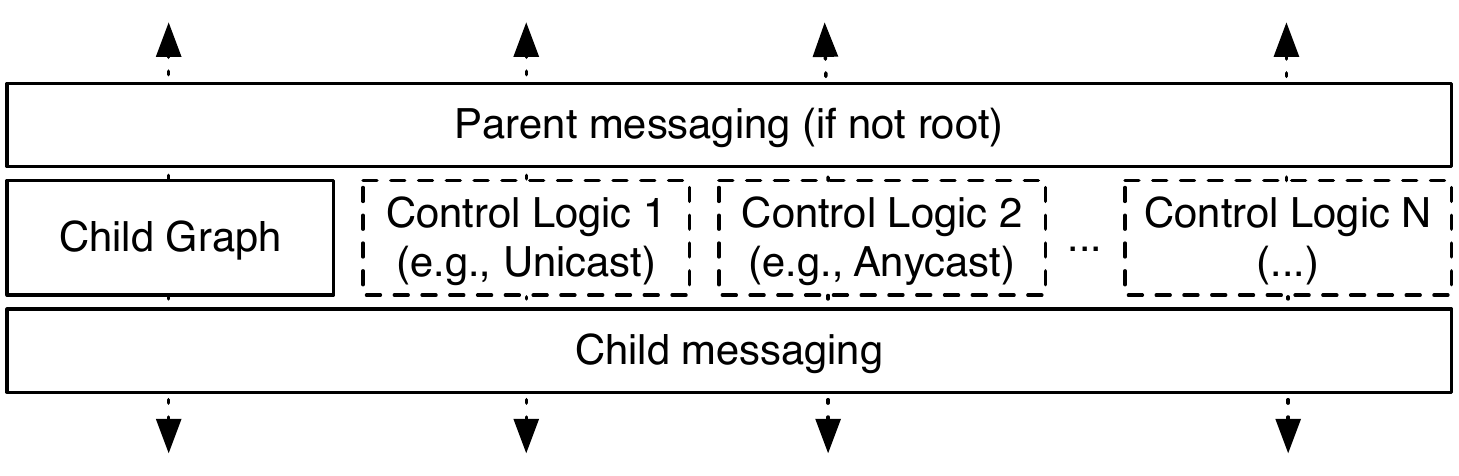}
\caption{Software structure in a normal (non-leaf) \lxb.  Solid outlines represent parts of the framework, dashed lines represent ``user'' provided control logic.
Leaf LXBs are similar, but include repair functionality and the bottom interface is towards physical switches rather than child \lxbs.
}
\label{f:nonleaf}
\end{figure}

{\bf Programmatic API:} This hierarchy of logical controllers provides a recursive programming model (schematically depicted in Figure \ref{f:nonleaf}), where for upward-bound computations, each \lxb accepts state from its children, performs some local computation on this state, and then exports information to its parent.  Downward computations are similar, with each \lxb accepting state from its parent, performing some local computation, and then pushing state down to its children. The nature of the state being pushed up or down and the nature of the local computation are fully general,
though a scalable control logic design is likely to make state less detailed as it flows up the hierarchy (\eg rather than simply pushing up a list of all failed links, push up only information on failed paths that cannot be recovered locally), and more concrete and detailed as it flows down the hierarchy (\eg combining high-level forwarding information from the parent with local knowledge of the topology to create forwarding information for individual children).
Moreover, computations need not be strictly upwards or downwards, but can push information in both directions as needed.
Note that it is trivial to achieve locality-of-control in such a recursive structure.  For example, the recomputation of routes in response to a failure is easily restricted to the affected region: if the upwards computation at a tier results in no change to the state it previously sent its parent, then the computation need not proceed further upward.

More specifically, the API presented to control logics consists primarily of three things: information about the corresponding \lxb (\eg its unique ID, which tier it is on), information about its local neighborhood (\eg its parent, a graph of its children), and mechanisms for sending and receiving messages to and from the parent and children controllers. At a conceptual level, RSDN control logics are built around two basic event handlers: ParentMessage and ChildMessage. These are invoked whenever an \lxb receives a message from a parent or child node, and initiate some internal computation and then optionally result in messages being sent to either parent or child nodes. New control logic designs are specified by these two event handlers, which can be used (as we discuss later) to build a variety of routing and traffic engineering solutions.

While we do not describe the RSDN implementation in detail, we note that control logics require special care at leaf \lxbs, as the children are hardware switches and not other \lxbs. More generally, while we describe the system as running the same exact control logic code in each \lxb (and this is the case for the logics we discuss in detail in subsequent sections), there may be cases where doing otherwise has benefits.  For example, one might use different routing within datacenters than between them.  This is entirely possible: as long as control logics implement the same interface (\eg send the right messages to their parents and understand the right messages from their children), they can be freely mixed.

{\bf Network repair:} In addition to providing an API to control logics, the RSDN framework also provides a network repair mechanism. This mechanism locally directs traffic around failures (independent of the routing computation itself), by making novel use of network virtualization: the repair mechanism implements a virtual ``failure-free" network on the physical network (which contains failures).

{\bf Integration with rest of the network:}
RSDN focuses on the core routing functions for carrier networks, but it can cleanly coexist with other aspects of carrier control planes.  As described in \cite{fabric}, we assume the network edge handles the ``host-network'' interface, while RSDN merely delivers packets from ingress port to the appropriate egress port as determined by the edge (that is, RSDN implements a \emph{fabric} in the terminology of \cite{fabric}). The edge control structure -- which could involve SDN controllers or RCP-like BGP implementations \cite{rcp} or even standard BGP/iBGP implementations (with RSDN functioning as a BGP-free core) -- provides the state to the edge routers,\footnote{The edge routers need not be physically distinct from the first-hop RSDN routers, but could refer instead to a separate set of tables that is processed first on ingress and processed last on egress.} and these edge routers (i) map incoming packets to the appropriate RSDN egress ports and (ii) may insert MPLS-like headers that tell the RSDN-controlled routers to which egress port the packet should be delivered (such additional headers would be stripped off at egress).

Similarly, the edge control structure can insert ACLs into edge routers as needed, or direct packets through tunnels (for network virtualization) or middleboxes (located at the edge) before entering the RSDN-controlled core. Thus, RSDN is merely providing a global-scale fabric around which additional functionality can be inserted at the edge in a straightforward manner.

\section{Context}
\label{sec:context}

{\bf Related Work:} Following our own initial work on the subject\cite{ons}, there have been two sets of short, related papers (\cite{iris, raon1, raon2}, \cite{cc1, cc2} ) some referencing our work, some not.  None of these papers (our own initial work included) explore the framework nature of RSDN, provide sufficient flexibility to implement effective routing at large scale, look beyond unicast routing, or provide general repair mechanisms (all of which we do here).

More generally, there is a vast literature on routing and related topics.
While the widely-deployed OSPF and IS-IS protocols utilize two levels of hierarchy to scale, perhaps the most relevant entry in the hierarchical routing literature is PNNI.
PNNI \cite{pnni} is a routing solution developed for ATM which, like RSDN, aggregates switching hierarchically.  However, it does not provide a framework for hierarchical computation of the various aspects of routing (\eg multicast, traffic engineering). That is, PNNI is hierarchical, but not programmably recursive -- it has no  programmatic interface that allows one to run the same control logic at each level of the hierarchy.  Moreover, PNNI's relationship to ATM limits it to operating on virtual circuits.  Thus, while PNNI's model of hierarchically aggregated switching blazed many of the relevant trails, it cannot supply the general and scalable programmability that global-scale SDN-based carrier networks need.

In terms of applying SDN to global-scale networks, B4 is the most relevant work \cite{b4}. B4 is a sophisticated traffic engineering solution for a network that interconnects a few dozen Google datacenters, and we see it as a brilliant solution to a different problem. As a routing control plane, we note that B4 copes with geographic scope, but not numerical scale. As an exercise in traffic engineering, B4 leverages the ability to control its own traffic at the edge, and give some traffic low-quality service, to achieve extremely high utilizations; in the traffic engineering designs we present here (which more closely represent current carrier requirements), we do not assume one can throttle edge traffic nor that there are low-quality classes of service one can use to keep utilization high.

RSDN's network repair algorithm is a generalization of local protection and an application of network virtualization, and thus has roots in and similarities to previous literature (\eg MPLS FRR~\cite{rfc4090,rfc6981}), but we are not aware of any existing work that combines them in the way we do here.  There is also a conceptual similarity to consensus routing\cite{consensus} in that the safety and liveness of packet delivery are distinctly separated.  While consensus routing separated these along the lines of consistency (\ie safety was associated with stable, consistent routing state and liveness with temporary but possibly inconsistent state to handle failures), RSDN's state might always be considered consistent by adapting techniques from the general SDN literature~\cite{consistentupdate,incrementalconsistentupdate}.  In RSDN, one might consider paths incorporating TE as safe, with temporary (but TE-unaware) state established by RSDN's repair algorithm in response to failures as providing liveness.

{\bf Topologies:}
\label{sec:topo}
To evaluate RSDN, we performed simulations on synthetically generated graphs.  We created our own topology generator, rather than taking advantage of one of the many existing generators (\eg \cite{brite,gtitm}) or measured real-world Internet topologies (\eg \cite{Spring02measuringisp,discarte}), because none of the resulting topologies matched the high level description we heard in our conversations with carriers.  Li et al.\cite{doyle} discuss likely reasons for the disconnect between these generated/measured topologies and the real world, which include the questionable realism of degree-distribution generation approaches and the presence of MPLS confounding IP-based mapping techniques.

Our generator's basic unit is a country (or continental) network.  Within each country, there are multiple metropolitan area networks, and each metro network has some number of access networks.  The metro networks within a country are connected by a core backbone network.  Multiple countries are interconnected to create a global network.  Metro networks are connected to cores redundantly such that no single switch or link failure can disconnect a metro from its core.  Country cores are interconnected by placing the countries on a 2-D plane and connecting nearby ones, similar to \cite{waxman}. We use a simple distance model where links between metros are ten times longer than within metros, and links between countries are twice the length of links between metros.

At a high level, our generator is consistent with the description of ``heuristically optimal'' networks described in \cite{doyle}.  While both \cite{doyle} and discussions with carriers indicate the existence of very large tree-like access networks, we actually disable their generation for all of the topologies used in this paper, and we consider the points at which the access networks would attach as the edge of the RSDN network.  The rationale here is simple; in trees, route computation is trivial (or even unnecessary) and traffic engineering and failure recovery are impossible, so extending our simulations to access networks would only belabor the obvious (and artificially make our results look better by trivially achieving optimal results in this part of the network).

Our generator has many parameters that control, for example, the number of nodes and the degree of connectivity within cores and within metros, the number of countries and how many connections there are between neighboring countries, and so on.  In our evaluations, we run our simulations on several graphs that explore a range of parameter settings.  All of our simulations were run on three-tier topologies, but this is a property of our topology generator and not a limitation of RSDN, which does not enforce that the hierarchy be any particular depth or even that it be uniform.

{\bf What Matters:}
One considers issues like route computation time, path stretch, and routing state when considering a routing algorithm. However, RSDN supports a wide variety of route computation algorithms offering a range of tradeoffs between these metrics. Thus, while these quantities are important for evaluating whether a particular route computation algorithm is suitable for a given network, they are not properties of RSDN itself. Thus, what really matters about RSDN is whether: (i) RSDN enables a broad enough class of routing algorithms (broadly defined) to meet various needs and (ii) RSDN-controlled networks can respond quickly enough to failures to achieve high availability. The first issue is addressed by our description and evaluation of two unicast routing schemes (Section \ref{sec:unicast}) and two traffic engineering schemes (Section \ref{sec:te}), as well as brief descriptions of anycast and multicast routing (Section \ref{sec:misc}). The second issue is addressed by our inclusion of a network repair mechanism in RSDN (Section \ref{sec:repair}), which provides -- independent of any routing algorithm -- a rapid failure recovery mechanism.


\section{Unicast Routing}
\label{sec:unicast}

The design space for unicast routing solutions is vast.  The goal of RSDN is not to pick one particular approach, but to enable operators to choose or develop one that is suited for their needs while leveraging the recursive structure of RSDN to scale. To illustrate how RSDN supports route computation, we discuss and evaluate two example routing control logics (out of the many that we have implemented and experimented with), both of which are implemented using an up-phase and then a down-phase.

\subsection{Fine-Grained Routing (FGR)}
\label{sec:fgr}
\begin{figure}[t]
\centering
\includegraphics[width=3.09in]{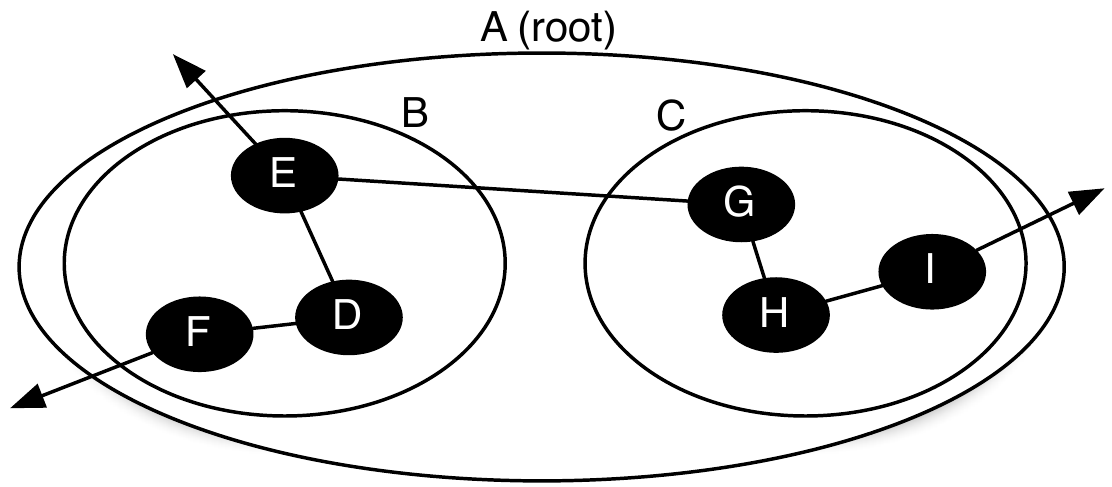}
\caption{A two-tier hierarchy where the root \lxb ($A$) has two children, $B$ and $C$.  Physical switches are shown in black; higher-level \lxbs are shown in white.  $E$, $F$, and $I$ are top-tier border switches, as they have links which cross the top-tier (root) \lxb.
}
\label{f:fgr}
\vskip -1em
\end{figure}

First we consider Fine-Grained Routing, in which
we compute shortest paths across an entire RSDN network.\footnote{There is one caveat to this claim of optimality: we explicitly disallow paths that must exit and later re-enter the same \lxb.  If the only optimal paths violate this rule, we will choose a suboptimal path, but these cases did not arise in our topologies.}  The problem is broken down recursively: each \lxb's controller is tasked with computing paths between the \lxb's \emph{border switches}, which are simply the physical switches within \anlxb that connect to something outside the \lxb~ -- often to sibling \lxbs, but also possibly to the outside world.  Examining Figure~\ref{f:fgr}, $F$, $E$, and $I$ are border switches of \lxb $A$; $E$ and $F$ are borders of $B$; and $G$ and $I$ are borders of $C$.

At the start of the upward pass of this algorithm, the controllers of the bottom-most \lxbs compute shortest paths between each of their border switches.  For example, $B$ would compute a path between $E$ and $F$.  A distance matrix containing the distances between all of $B$'s border switches is pushed upwards to its parent on the next higher tier -- in this case, $A$.  Once $A$ has border-to-border distance matrices from each of its children, it computes shortest paths between each of its own border switches --- in this case, paths between all combinations of $E$, $F$, and $I$.

In the downward pass of this algorithm, the parents use the paths they computed on the upward phase to push down actual forwarding rules to forward between their children.  These rules accumulate down the hierarchy.  For example, $A$ sends rules to $E$ so that it can reach $G$ and vice-versa, and $B$ sends rules to forward between $F$, $D$, and $E$.  The aggregation of all such rules when they reach the switches at the bottom of the hierarchy is sufficient to communicate between all borders of an LXB -- most crucially, between all the border switches of the root \lxb, since these are the switches that connect to the outside world.

Note that $A$ does not need to compute paths between switches which are not its own borders (\ie $D$, $G$, and $H$).  Nor does $A$ know anything at all about switches which are not borders of its children; $H$, for example, does not appear in the distance matrix provided by $C$ and is therefore invisible to $A$.  Also note that this means that a given \lxb's borders are always a subset of the union of the border switches of its children.  Finally note that siblings (\eg $B$ and $C$) need not know anything about each other -- only the common parent need know something about both -- and therefore can compute entirely in parallel.  This combination of information hiding and parallelization allows FGR to scale considerably better than a flat routing computation, and the information hiding also reduces the amount of state that must be shared between controllers (and ultimately placed on switches).  However, as the number of border switches increases, these advantages lessen, possibly necessitating another solution such as the one we explore next.

\subsection{Coarse-Grained Routing (CGR)}
\label{sec:cgr}
\begin{figure}[t]
\centering
\includegraphics[width=3.09in]{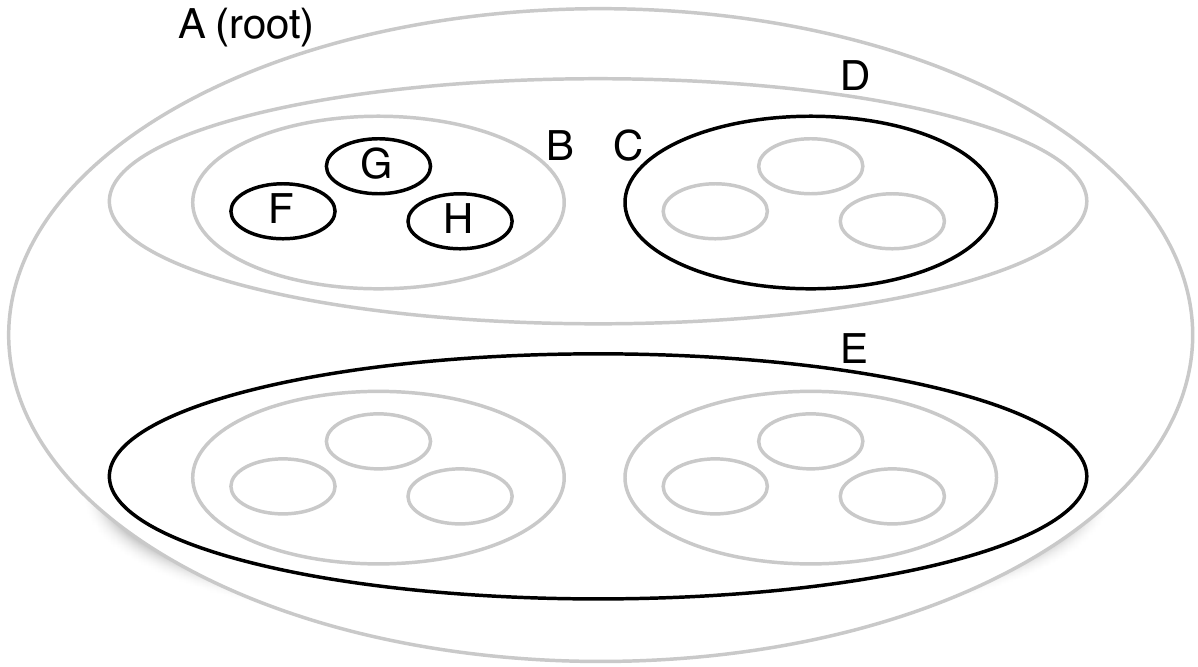}
\caption{A three-tier hierarchy highlighting the \lxbs for which switches $F$, $G$ and $H$ have forwarding entries.  Links have been omitted for clarity.  Unlike Figure~\ref{f:fgr}, physical switches are not colored differently as they can be considered as just another \lxb in CGR.
}
\label{f:cgr}
\vskip -1em
\end{figure}

To achieve better scalability, we have developed a coarser-grained approach to routing where a switch need only contain forwarding state to reach sibling \lxbs.  For example, Figure~\ref{f:cgr} emphasizes the \lxbs for which switches $F$, $G$, and $H$ must contain forwarding state.  First, they must hold state about each other, as they are all siblings -- they share parent $B$.  Second, they must hold forwarding state for $C$, as $B$ and $C$ are both children of $D$.  Third, they must hold forwarding state for $E$ as it shares a parent ($A$) with $D$. This approach results in an enormous savings in both state and computation relative to FGR because it nests routing and hides significant information about the network.  The natural downside to this information hiding is that the ``further away'' a switch is, the more inexact forwarding entries to it are, and thus path length may increase.

A controller running CGR computes a number of \emph{possible} paths between its children, and informs the children of all of these possibilities.  This tells the child that it can reach some sibling $X$ with distance $Y$ via an external link $Z$.  These possibilities are passed down the hierarchy with each tier pruning ones which are clearly not optimal (\eg when the distance of one path plus the diameter of the child is less than the distance of a second path, the second route could never be better).  Upon reaching the leaf \lxbs at the bottom of the hierarchy, the best of the remaining paths to an \lxb is in fact the optimal path to the closest point of that \lxb.  That is, $F$ could find an optimal path to $G$ and $H$, as well as an optimal path to whatever switches in $C$ and $E$ are closest to $F$.  In these latter cases, a packet that reaches the destination \lxb would then get an optimal path to the next destination \lxb one tier down the hierarchy, and so on recursively until reaching the bottom tier of the hierarchy -- the physical switch which is the destination.  Thus, one can view this as a greedy recursive shortest path algorithm.

\subsection{Evaluation of Routing Algorithms}
\label{sec:routingevaluation}

\begin{table}[t]
\centering
\begin{tabular}{|l|l|l|l|l|l|l|l|l|}
\hline
\textbf{Switches} & \textbf{Links} & \multicolumn{3}{|c|}{\textbf{Children/Tier}}    & \textbf{Avg. Link} \\\cline{3-5}
                   &                 &  1 &  2 &  3 &  \textbf{Latency} \\\hline\hline
10355    & 47595     & 19 & 665 & 10355 & 14.584ms                      \\\hline
\end{tabular}
\caption[Topology Characteristics]{Characteristics of Topology X used for our unicast routing experiments.}
\label{table:topo_chars}
\vskip -1em
\end{table}

We presented FGR and CGR merely to demonstrate that RSDN could incorporate widely varying routing designs, and there are many other routing designs one could devise, but here we discuss how FGR and CGR perform in terms of stretch, computation time, and routing table size. For clarity, our presentation focuses on results from a single topology (call it topology X) that is described in Table \ref{table:topo_chars}. However, we also performed extensive experiments with nine additional topologies with between $1,000$ and $10,000$ switches, and degrees of intra- and inter-country connectivity spanning over a factor of three (ranging both above and below what we believe to be realistic). The general conclusions we draw here about RSDN's performance are consistent with the results from those additional experiments.

\begin{table}[t]
\centering
\begin{tabular}{|l|p{0.18\columnwidth}|p{0.18\columnwidth}|p{0.18\columnwidth}|}\hline
\textbf{Tier} & \textbf{CGR} & \textbf{FGR} & \textbf{APSP}\\\hline
3 & \hfill$0.407$s & \hfill$0.391$s & ---\\\hline
2 & \hfill$5.709$s & \hfill$6.036$s & ---\\\hline
1 & \hfill$6.745$s & \hfill$6.811$s & ---\\\hline
0 (Root) & \hfill$7.215$s & \hfill$249.081$s & \hfill$682.452$s\\\hline
\end{tabular}
\caption{Times to run computation on topology X up to given level in the hierarchy using FGR, CGR, and running all-pairs shortest path (APSP) over the entire topology.}
\label{t:computation}
\vskip -1em
\end{table}

\begin{figure}[t]
\centering
\includegraphics[width=3.23in]{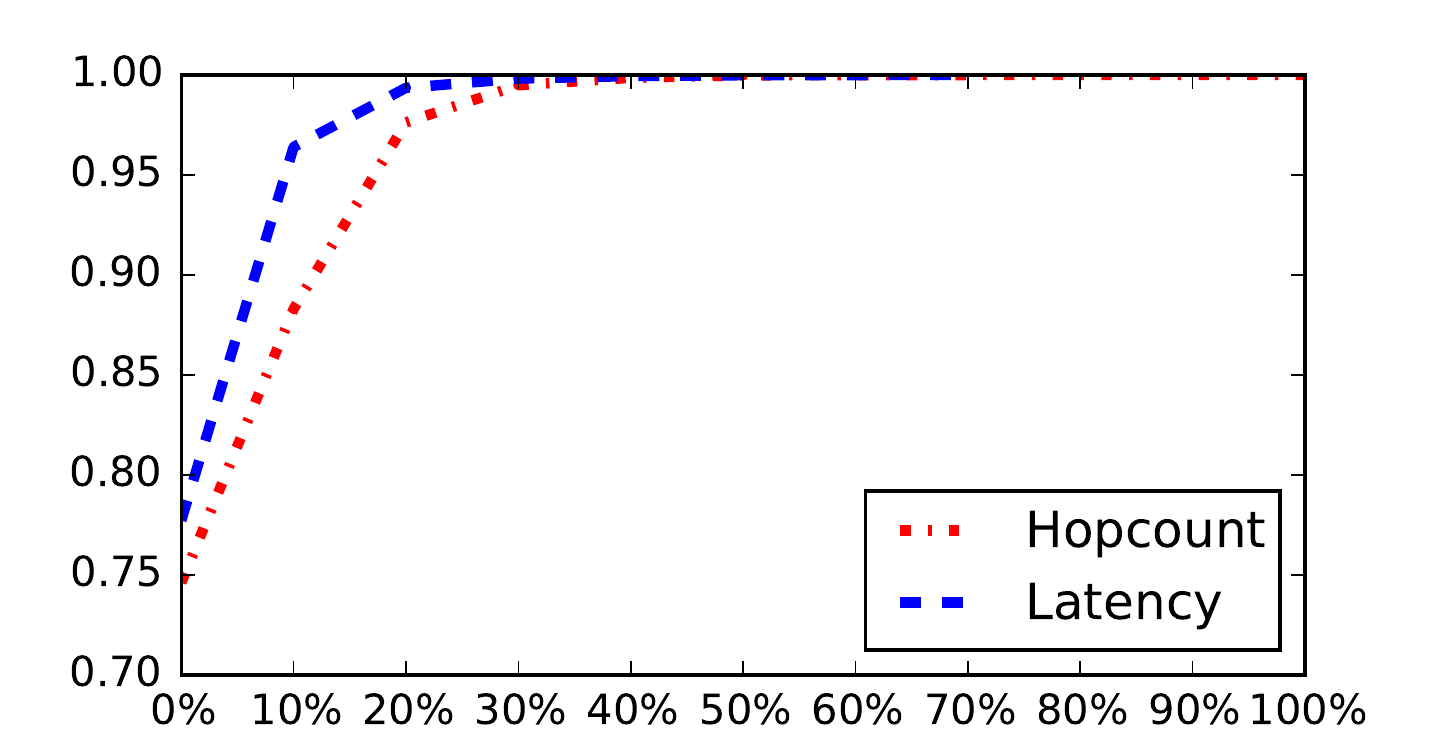}
\caption{The CDF of network stretch (as fractional increase over original path) for CGR on topology X.}
\label{f:stretch-with-cgr}
\end{figure}

{\bf Computation Time:} We compared our Python RSDN route computations with a Python All-Pairs-Shortest-Path (APSP) computation. The numbers here are not indicative of how an optimized C++ computation would perform, but the relative performance provides some measure of the underlying computational complexities. Route computation times are shown in Table~\ref{t:computation} for FGR and CGR up to each tier (this is relevant because after an initial computation, one only needs to recompute to establish good routes after a failure; in only very few cases -- a failure of a country-to-country link in our topologies -- does this require recomputing all the way up to the root). Note that RSDN partially parallelizes the computation (since \lxbs in the same tier can compute in parallel), which is why FGR beats APSP in computation time even though they both compute globally shortest paths.  While CGR does not compute shortest paths, it trades this off for significantly faster computational performance.

{\bf Stretch:} FGR computes exact shortest path routes, so there is no stretch. The stretch induced by CGR on topology X is summarized in Figure \ref{f:stretch-with-cgr}, which shows the CDF of stretch over all source-destination pairs using both hopcount-based routing and latency-based routing.\footnote{Latency-based routing minimizes path length in terms of distances, and thus takes into account the geography as determined by our topology generator.}  For hopcounts, fewer than 15\% of the pairs have stretch over 10\%; for latency, approximately 5\% do. These latency results were quite consistent with the average of all nine additional topologies we tested: 75\% of the paths have no additional stretch and 92\% of the paths have have less than 10\% stretch.  For hopcount-based routing on these additional topologies, the results were somewhat worse: 70\% of the paths have no additional stretch, 73\% of the paths have have less than 10\% stretch, though 93\% of the paths have less than 20\% stretch.  Intuitively, hop counts may become worse, though latency does not suffer as much because CGR gets the broad strokes right (routing to the correct geographical region at some scale -- accounting for the majority of latency) but may get the details wrong (routing to a suboptimal switch within the region -- corrected by additional lower-latency hops).
Ultimately, CGR trades off moderate stretch for significantly faster computation.

\begin{table}[t]
\centering
\begin{tabular}{|l|p{0.195\columnwidth}|p{0.195\columnwidth}|p{0.195\columnwidth}|}\hline
\textbf{Routing} & \textbf{Label} & \textbf{LPM Cons.} & \textbf{LPM Rand.}\\\hline
 CGR & $211.56$ & $382.09$ & $13,088.38$\\\hline
 FGR & $2,942.76$ & $1,296.02$ & $41,206.13$\\\hline
\end{tabular}
\caption{Average table size with labels, highly-aggregatable (consecutive) IP prefixes, and poorly-aggregatable (random) IP prefixes. }
\label{t:aggregation}
\end{table}

{\bf Table Size:} RSDN can incorporate a variety of approaches to forwarding tables and address assignments, and here we discuss two of them. First, we considered an MPLS-like scheme where at the network edge one or more labels are applied to the packet, and all forwarding is done on these labels. Table \ref{t:aggregation} shows the resulting average forwarding table sizes that arise using this approach for both FGR and CGR on topology X, with CGR being significantly smaller than FGR (about 93\% savings).  We then tried using IP addresses for forwarding (with LPM aggregation in the tables) considering two address allocation schemes. The first took a set of 495k prefixes from Route Views \cite{routeviews} and assigned them consecutively to the exterior ports (putting the same number of prefixes on each port); the second assigned these randomly to the exterior ports.  Unsurprisingly,
the easily aggregated consecutive assignment yielded smaller tables, though CGR provided about 70\% savings for both.
These results show that: (a) RSDN can use either labels or aggregatable addresses, and (b) if small table size is important, CGR provides an effective and relatively low-stretch way of accomplishing this.


\begin{table*}[!t]
\centering
\begin{tabular}{|l|l|l|l|l|l|l|l|l|}
\hline
\textbf{Topo} & \textbf{Switches} & \textbf{Links} & \multicolumn{3}{|c|}{\textbf{Links}} & \multicolumn{3}{|c|}{\textbf{Children}} \\\cline{4-9}
&                 &                 & Tier 1 & Tier 2 & Tier 3 & Tier 1 & Tier 2 & Tier 3 \\\hline\hline
little-low  & 119 & 224 & 28 & 42 & 154 & 7 & 28 & 119          \\
little-med  & 119 & 230 & 30 & 42 & 158 & 7 & 28 & 119          \\
little-high & 119 & 252 & 33 & 42 & 177 & 7 & 28 & 119          \\\hline
middle-low & 170 & 303 & 27 & 60 & 216 & 10 & 40 & 170         \\
middle-med & 170 & 343 & 31 & 60 & 252 & 10 & 40 & 170         \\
middle-high & 170 & 350 & 37 & 60 & 253 & 10 & 40 & 170         \\\hline
big-low & 255 & 459 & 55 & 90 & 314 & 15 & 60 & 255         \\
big-med & 255 & 500 & 62 & 90 & 348 & 15 & 60 & 255         \\
big-high & 255 & 547 & 68 & 90 & 389 & 15 & 60 & 255         \\\hline
\end{tabular}
\caption[Topology Characteristics]{Characteristics of topologies used for our TE simulations.  These are small relative to those we use to evaluate routing due to the fact that we cannot compute the gold-standard algorithm in reasonable times on larger graphs. }
\label{table:tetopos}
\vskip -1em
\end{table*}
\section{Traffic Engineering}
\label{sec:te}

\newcommand{\RLP}{RLP\xspace}
\newcommand{\DP}{RST\xspace}

As with our exploration of unicast routing, our goal here is not to pioneer radically new TE paradigms but instead to merely show that RSDN can scalably achieve TE goals. There are many TE designs, but most of them can be grouped into two categories.  In the first, the traffic matrix is fed into an offline solver that generates optimal routes.  The second is used in conjunction with multipath routing: a feedback system relays congestion information about a path to its source, and the source preferentially steers traffic over less-loaded paths. We have implemented recursive versions of both approaches, which we term Recursive Linear Programming (\RLP) and Recursive Split Tuning (\DP).

For our evaluation, we focus purely on whether good routes can be chosen, and ignore real-world issues such as flow divisibility and packet reordering (which are issues for any TE design). The primary metric by which we evaluate our TE approaches is the load of the maximally loaded link (and by load we mean utilization level), which should be minimized.  We compare both of our RSDN TE implementations against a ``gold standard'' -- a straightforward global linear program that achieves the minimum possible value of this metric.

\subsection{Recursive Linear Programming}
A simple approach for a recursive linear program would be to take the gold-standard linear program (minimizing the maximal load) and run it in each LXB using the graph of its child LXBs and the links between them.  Such an approach is nicely recursive: the root LXB performs traffic engineering at the coarsest level, and each lower LXB is responsible for a less abstracted and more localized part of the problem than its parent.  However, this approach is ineffective in practice: it hides too much information, as a parent LXB does not know if additional load on a child will increase its maximal load.

We could address this by exposing more information, but instead we chose an iterative approach where the parent learns about the effect of additional load from the previous iteration, rather than being given enough information to have figured it out a priori. Each child in our \RLP implementation only reports a single value to its parent: the load of the most-loaded link between or within any of its own children. We extend the basic LP to compute paths that not only minimize link loads, but also node loads (which are the reported loads from child \lxbs).

\subsection{Recursive Split Tuning}

The second approach to TE uses multipath routing and adaptively and iteratively balances traffic across the available paths at the source or the point of ingress using some type of path-load feedback mechanism, as in \cite{mate,texcp,date,replex}.  This requires RSDN to implement the various pieces of this approach: (i) a multipath routing algorithm, (ii) a mechanism to adjust the balance of traffic across various paths, and (iii) a method for gathering information about the load along a path.

We experimented with several methods for choosing multiple paths, but the most straightforward -- and the one we evaluate here -- is a simple variation of the fine-grained routing algorithm discussed in Section~\ref{sec:fgr}.  Rather than pick a single shortest path, we use Suurballe's algorithm\cite{suurballe} to choose two disjoint paths (if possible).   The iterative algorithm that actually determines the split values is simple: when one of the two possible paths is more loaded than the other, the split is adjusted to favor the less-loaded link slightly more.  The algorithm to actually find the load of a path is slightly more complex -- as with \RLP, the relevant information (the most loaded link along a path in this case) may be hidden within a child.  To confront this, each child informs its parent of the load of the most-loaded link being used between every pair of ports.

That this same balancing algorithm is run in each LXB, and thus at each tier of the hierarchy, has two related implications worth pointing out.  First, while two paths may seem insufficient (some of the cited prior work uses many more), two paths are chosen \emph{for every \lxb}.  If we consider a source-destination pair between edge ports, which is balanced across two paths, and each of those paths goes through two child LXBs, this yields eight different paths with only two tiers of hierarchy; more tiers or traversing additional LXBs quickly grows the path count.  Second, when run at a lower tier, the algorithm makes more localized changes than when run at a higher tier.  This motivates having lower tiers iterate with a higher frequency than higher tiers; this way, there is a chance to make small (more localized) adaptations before making larger ones (we arbitrarily chose a ratio of five to one between the frequency of each tier and the one above it).

\subsection{Evaluation of Traffic Engineering}

\begin{table}[t]
\centering
\begin{tabular}{| l | l | l | l | l |}
\hline
 \textbf{Topology} & \textbf{RLP} & \textbf{RST} & \textbf{None} & \textbf{G-S}  \\
 \hline\hline
 little-low & 1.00 & 1.00 & 1.48 & 1.00  \\
 little-med & 1.00 & 1.00 & 1.60 & 1.00 \\
 little-high & 1.00 & 1.00 & 1.20 & 1.00 \\
 \hline
 middle-low & 1.00 & 1.22 & 2.22 & 1.00 \\
 middle-med & 1.00 & 1.00 & 1.58 & 1.00 \\
 middle-high & 1.00 & 1.00 & 1.84 & 1.00 \\
 \hline
 big-low & 1.00 & 1.00 & 1.79 & 1.00 \\
 big-med & 1.00 & 1.00 & 1.56 & 1.00 \\
 big-high & 1.00 & 1.00 & 1.87 & 1.00 \\
 \hline
\end{tabular}
\caption{Maximal link load normalized by that achieved by the globally optimal gold-standard on nine topologies.}
\label{table:te_max}
\vskip -0.5em
\end{table}

\begin{figure*}[t]
\centering
\subfigure[][]{
   \includegraphics[width=0.4\textwidth]{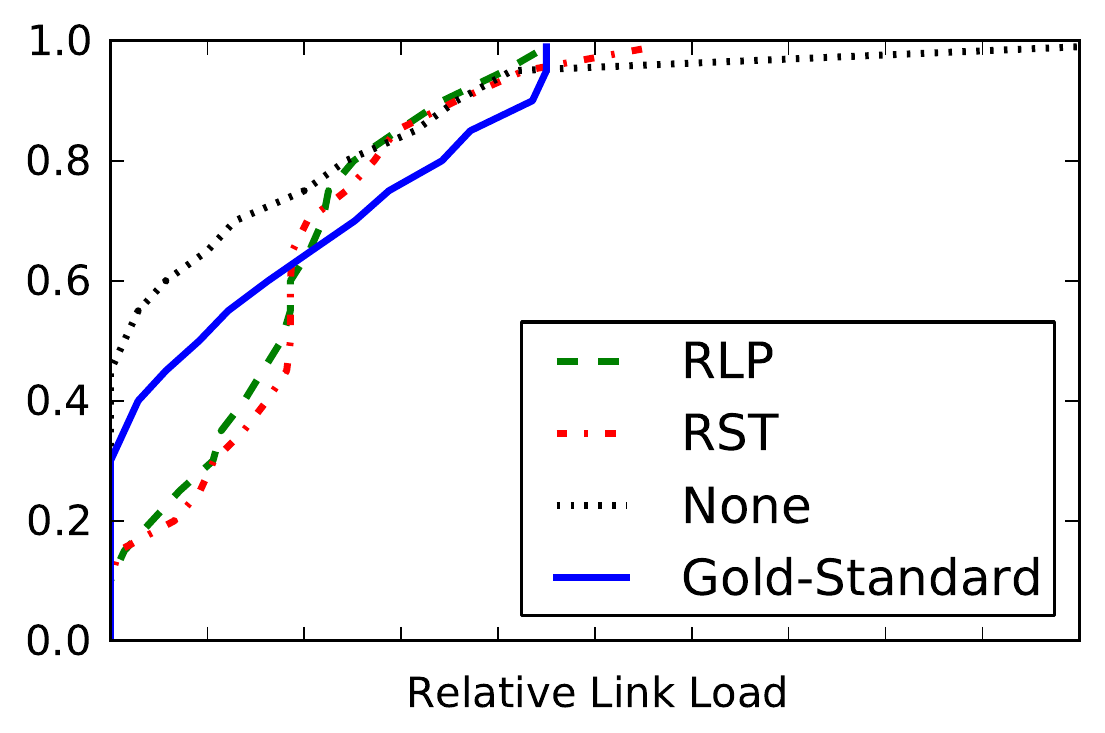}
   \label{fig:te_cdf1}
}
\hfill
\subfigure[][]{
   \includegraphics[width=0.4\textwidth]{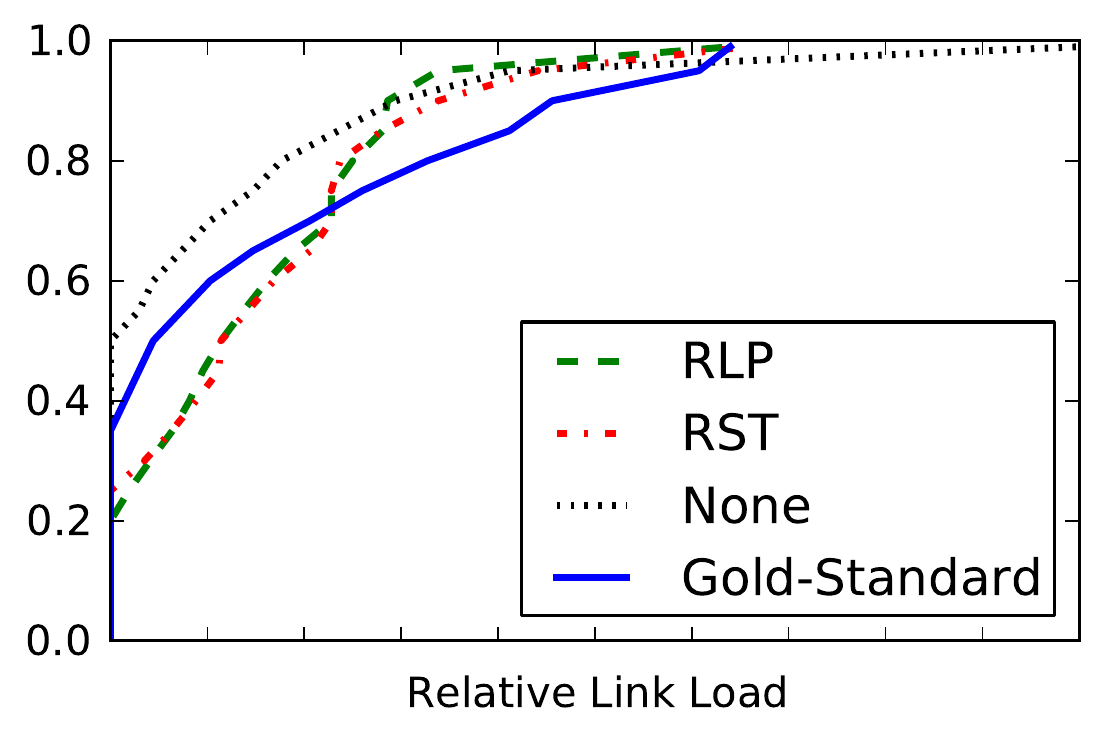}
   \label{fig:te_cdf2}
}
\vspace{-0.3cm}
\caption{
CDF of the load on links in the \subref{fig:te_cdf1} middle-low and \subref{fig:te_cdf2} big-med topologies when using each of the traffic engineering methods.
}
\label{f:te_cdf}
\vskip -1em
\end{figure*}

We evaluate TE by running simulations on nine topologies (characterized in Table~\ref{table:tetopos}).  Table~\ref{table:te_max} shows the worst-case link load for RLP and RST and compares them to the results from the global optimal (against which all the results are normalized) and from merely using shortest-path routing with no TE. In all cases, RLP matches the performance of optimal to within less than a percent.
RST is also generally within a percent of optimal; a single case misses by 22\%.
Additionally, leveraging locality means the recursive approaches converge faster: on average, RLP reaches its best case in 65\% of the time needed by the gold standard, and RST in only 15\%.

In Figure~\ref{f:te_cdf} we look at the distribution of link loads on two topologies. While the goal of the gold-standard TE algorithm is only to minimize the maximal load, the RLP and RST algorithms do a better job of spreading the load around (even if they cannot always achieve the same mini-max load). That is, RLP and RST have fewer highly loaded links (as compared to the gold-standard), having pushed some of that load off to the more lightly loaded links; the gold-standard does not bother trying to decrease loads on less-than-maximally-loaded links.\footnote{This is not to say that this sort of load spreading could not be made an objective of a global solver -- simply that it comes ``for free'' when using RSDN's recursive TE approach.}

While we admittedly ignore some real-world concerns in our simulations and analysis, the result we find is encouraging: we were able to take the two dominant TE paradigms and create recursive examples of each which fit nicely in the RSDN framework and perform competitively.


\section{Network Repair}
\label{sec:repair}

While the previous two sections have examined control logic written using the RSDN framework, this section examines a feature of the framework itself, which all control logics automatically benefit from: network repair.

\subsection{Motivation and Design}

A common practice for improving network availability is to implement link protection, in which for every link between two routers (or nodes, in the text below) \texttt{a} and \texttt{b}, an alternate path (not including that link) is precomputed and then immediately used as a failover route when the link goes down.  This works as long as the failover remains up, but cannot cope when multiple failures knock out both the primary and failover paths. RSDN uses link protection, but then adds a more general network repair mechanism that can recover from {\em all} failure scenarios (as long as a path exists).

Our network repair approach is inspired by network virtualization.  When routes are computed, we note the state of the network (\ie which nodes and links are up).  We then embed a virtual version of this network within whatever the current physical network happens to be; this virtual network clearly supports the previously computed routes, so they need not be recomputed.  Note that RSDN's repair approach does not rely on the recursive structure and could therefore be implemented on any network.

\begin{table}[t]
\small
\centering
\begin{tabular}{|p{1.0in}|p{1.75in}|}\hline
\verb=Nbr[a]= & Neighbors of \verb=a=\\\hline
\verb=HavePath= & Nodes to which \verb=a= already has a path.\\\hline
\verb=NeedPath= & Nodes to which \verb=a= needs a path.\\\hline
\verb=Vnodes= & Nodes that \verb=a= ``virtualizes'' (includes the forwarding table for)\\\hline
\verb=EnsurePath(a,b)= & Ensures that there is a path between \verb=a= and \verb=b= considering the current state of the network.  Returns \texttt{true} if the link between \verb=a= and \verb=b= is up, has working link protection, or if a repair path can be computed.\\\hline
\end{tabular}
\caption{Network repair sets and functions for node \texttt{a}.}
\label{t:repair-functions}
\vspace{-0.15in}
\end{table}

To understand our repair algorithm, consider a particular network state (in terms of which nodes and links are up and down), and the sets and functions shown in Table~\ref{t:repair-functions}; the network repair algorithm at node \texttt{a} is as given in Algorithm~\ref{alg:networkrepair}.
After applying this simple procedure for each node, every node \texttt{a} has a set of internal tables that can be used to virtually route {\em through} unreachable nodes. That is, suppose that if \texttt{a} were to send a packet destined for some node \texttt{x} its first two hops would ordinarily be nodes \texttt{b} and \texttt{c}.
When a link fails, if the responsible controller discovers that \texttt{a} can no longer reach \texttt{b} (because \texttt{EnsurePath(a,b)} fails), then the controller \emph{virtualizes} \texttt{b}  within \texttt{a} (by including \texttt{b}'s routing table in \texttt{a}), determines where \texttt{b} would have sent the packet (in this case, node \texttt{c}), and computes a repair path to \texttt{c} using \texttt{EnsurePath(a,c)}. 
If this succeeds, then \texttt{a} can forward the packet directly to \texttt{c}; if it fails, then the procedure recurses and \texttt{a} imports \texttt{c}'s routing table, determines where \texttt{c} would have routed the packet, and then attempts to directly route to that node.

This procedure is initiated whenever a neighbor fails (or otherwise becomes unreachable), and results in a complete set of routing tables.  The computational complexity of this operation scales not with the overall network size but with the complexity of computing repair paths between nearby nodes (\ie when \texttt{EnsurePath} finds that there is no direct or protected link between \texttt{a} and \texttt{b} and must compute a new path).  The maximal distance between any two nodes where this function is invoked is the most consecutive unreachable nodes in a path.  It is very unlikely that this distance will ever be more than a few hops.

\begin{algorithm}
\small
\centering
\begin{algorithmic}[0]
\caption{Network Repair Algorithm.}
\label{alg:networkrepair}
\LineComment{Repair algorithm for node \texttt{a}}
    \State \texttt{NeedPath} $\gets$ \texttt{Nbr[a]} \Comment{Initialize \texttt{NeedPath}}
    \State \texttt{HavePath} $\gets \emptyset$ \Comment{Initialize \texttt{HavePath}}
    \State \texttt{Vnodes} $\gets$ \texttt{\{ a \}} \Comment{Initialize \texttt{Vnodes}}

    \While{\texttt{NeedPath} $\neq \emptyset$}
        \State \texttt{b} $\gets$ \texttt{member(NeedPath)} \Comment{Take element \texttt{b}}
        \State \texttt{NeedPath} $\gets$ \texttt{NeedPath}$-$\texttt{b}
        \If{\texttt{EnsurePath(a,b)} $=$ \texttt{true}}
            \LineComment{If we can find a path from \texttt{a} to \texttt{b}, we're done}
            \State \texttt{HavePath} $\gets$ \texttt{HavePath} $\cup$ \texttt{b}
        \Else
            \LineComment{If no path is found, virtualize b}
            \State \texttt{Vnodes} $\gets$ \texttt{Vnodes} $\cup$ \texttt{b}
            \LineComment{Need paths to neighbors of b that we can't reach yet}
            \State \texttt{NewNbrs} $\gets$ \texttt{Nbr[b]}$-$\texttt{HavePath$-$\texttt{Vnodes}}
            \State \texttt{NeedPath} $\gets$ \texttt{NeedPath} $\cup$ \texttt{NewNbrs}
        \EndIf
    \EndWhile
\end{algorithmic}

\end{algorithm}

\subsection{Evaluation of Network Repair}

\begin{figure}[t]
\centering
\subfigure[]{
   \includegraphics[width=0.42\textwidth]{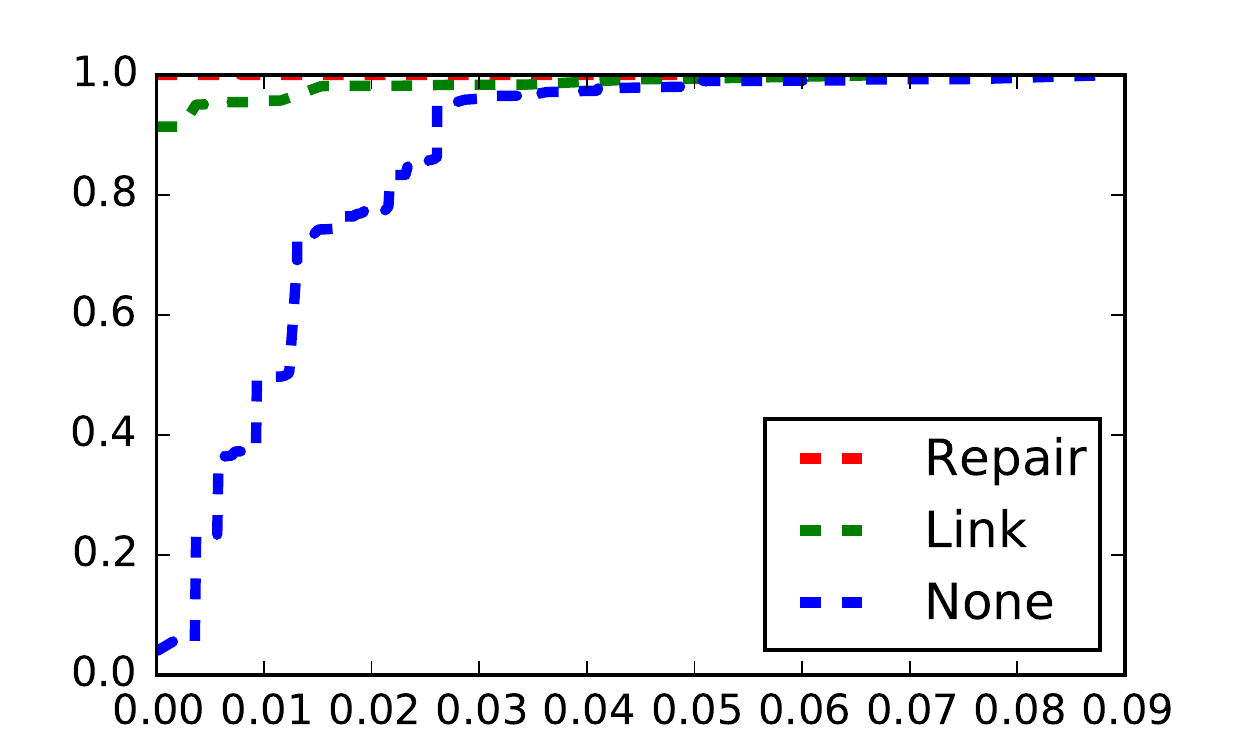}
   \label{fig:failure1}
}
\subfigure[]{
   \includegraphics[width=0.42\textwidth]{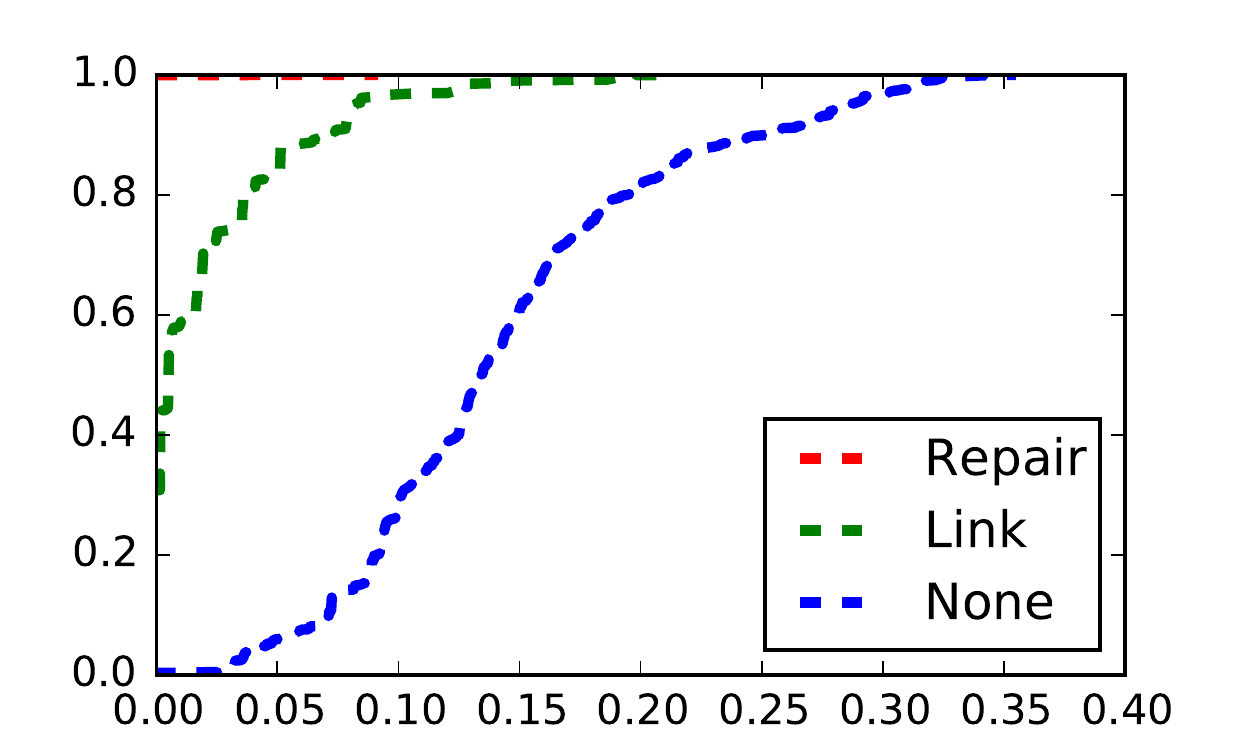}
   \label{fig:failure2}
}
\caption{CDF of CPRP (x-axis) for no failure recovery, link protection, and RSDN repair on the big-med topology under the \subref{fig:failure1} light and \subref{fig:failure2} heavy failure scenarios.
}
\label{f:availability}
\vspace{-0.12in}
\end{figure}

When considering failures, we ran experiments using a failure model based on \cite{sprint-failure-model}, which randomly fails and recovers links according to distributions measured in the Sprint network, and apply this same model to switches.\footnote{We do not include planned link outages, since those can be handled without loss of connectivity by most SDN-based systems.}  We looked at a variety of failure scenarios, but only report on two here: where links fail at an average rate of once per day, and where they fail once per ten days, and in each case we arbitrarily set the node failure rate to half of the link failure rate. 

Since no recovery mechanism can provide connectivity when the physical network is disconnected,
the appropriate performance metric is the percent of time physically connected pairs are connected by routing; or, for short, the connectivity of the physically reachable pairs (which we denote by \emph{CPRP}).\footnote{Note that it is a nongoal of network repair to perform traffic engineering on short time scales; that is, we accept the fact that when there is a failure, traffic may need to be routed in ways that unbalances the traffic. Indeed, we believe many carrier networks are run at relatively low utilization precisely so that these short-term traffic reroutings do not cause overload.}

\begin{table}[t]
\centering
\begin{tabular}{|l|p{0.22\columnwidth}|p{0.22\columnwidth}|p{0.22\columnwidth}|}\hline
 & \textbf{big-low} & \textbf{big-med} & \textbf{big-high} \\\hline
light &
$0.5$\%:$0.04$\% &
$0.5$\%:$0.04$\% &
$0.5$\%:$0.04$\% \\\hline

heavy &
$4.8$\%:$0.89$\% &
$4.2$\%:$0.9$\% &
$4.3$\%:$0.9$\% \\\hline

\end{tabular}
\caption{
Average link:node downtimes as percentages of the entire simulation runs for three different topologies.
}
\label{table:downtimes}
\vskip -1em
\end{table}
In analyzing the performance of network repair, we make the following two assumptions.  First, we only consider failures once they have been detected by the sending switch, since there is nothing a routing or recovery mechanism can do to deal with undetected failures. Second, we assume that the time to repair a failure is, on average, 50ms. This is based on estimates of how long it takes (i) for messages to travel from switch to controller and back (to which we allocate 10ms in each direction, which is roughly the average over all switch-to-controller latencies in our topologies; this number is low because most links are quite local to their bottom-tier controller), (ii) for the controller to compute repair paths and generate a response (to which we allocate 10ms, which is more than reasonable)\footnote{For all of the simulations, we also measured the time taken to compute the on-demand repair paths and found it to be minimal.  The mean time spent per failure is less than 0.5ms and the maximal time less than 10ms \emph{for all controllers combined} -- and this is for completely unoptimized Python code.}, and (iii) the time it takes to install new routes (to which we allocate 20ms). This last quantity is by far the most variable, as it depends on the number of routes, the router technology used, and other factors outside our control. Most importantly, this number can improve with better router technology, whereas the others are due to more inherent limitations.\footnote{In addition, insertion when doing exact matching (as in MPLS, which could be used here for internal forwarding) can be far faster than when inserting for LPM.}

We present performance results of network repair on the three big topologies from Section~\ref{sec:te}; we also ran experiments on all the other topologies from Section~\ref{sec:te} and omit the (similar) results for space. We considered three possible recovery strategies: (a) none, (b) only link protection, and (c) network repair (which includes link protection).  We then ran the simulation for the equivalent of two days, and recorded what fraction of paths between all source-destination pairs are connected by the three recovery strategies.  Table \ref{table:downtimes} shows the average link and node downtimes in these scenarios, which represent extreme tests of the repair system.

Table \ref{table:repairaverages} summarizes the CPRP for all three topologies under both failure scenarios, and 
Figure~\ref{f:availability} details the results in terms of the CDF of availability for the big-med topology.
Note that even under the heavy failure scenario, where links are down roughly 5\% of the time, network repair is able to provide ``five 9s'' of CPRP, while no protection offers no 9s and link protection offers a single 9. This high relative availability is because -- in addition to covering the exact same failures that link protection covers in exactly same way -- network repair can recover from \emph{all} failures (subject only to table size limits on the switch), and link protection simply cannot. The  only reason that repair does not achieve the maximum possible connectivity is due to the 50ms delay when controller involvement is required.

One might be surprised that such a mechanism can provide five 9s when the Internet is generally deemed to be less than three 9s. The distinction is that we are not counting the case where the network is physically disconnected (since neither routing nor repair can help there) and are only measuring the availability for the physically connected pairs.  What our results show is that with network repair, the routing algorithm is no longer the availability bottleneck, regardless of what routing algorithm you use. Instead, the availability bottleneck is strictly physical connectivity, which must be addressed by other means.

\begin{table}[t]
\centering
\begin{tabular}{|p{0.09\columnwidth}|l|p{0.11\columnwidth}|p{0.13\columnwidth}|p{0.23\columnwidth}|}\hline
\textbf{Sim.} & \textbf{Topo} & \textbf{None} & \textbf{Link} &\textbf{Repair } \\\hline
\multirow{3}{*}{light} & big-low & $97.8$\% & $99.79$\% & $99.999977$\% \\
 & big-med & $98.7$\% & $99.85$\% & $99.999982\%$ \\
 & big-high & $98.1$\% & $99.77$\% & $99.999983$\% \\\hline
\multirow{3}{*}{heavy} & big-low & $82$\% & $96.2$\% & $99.99979$\%  \\
 & big-med & $86$\% & $97.9$\% & $99.99983$\%  \\
 & big-high & $85$\% & $96.1$\% & $99.99982$\%  \\\hline
\end{tabular}
\caption{
Average CPRPs for no failure recovey, link protection, and RSDN repair on all three big topologies from Table~\ref{table:tetopos} under both the light and heavy failure scenarios.
\vspace{-2mm}
}
\label{table:repairaverages}
\end{table}


\section{Additional Issues}
\label{sec:misc}
While the previous three sections focused on the most important uses of RSDN (unicast routing and traffic engineering) and its most important built-in function (network repair), here we briefly discuss additional use cases (multicast and anycast) and additional built-in functionality (reliability, bootstrapping, security).

\subsection{Multicast and Anycast}

Here we sketch RSDN designs for multicast (which we have actually implemented and simulated) and anycast (which we have not).  Our goal here is merely to show that these services can be easily accommodated with the RSDN framework.

\paragraph{Multicast:} To implement multicast within RSDN, we have developed a CBT-like~\cite{cbt} approach that is similar to other hierarchical multicast approaches~\cite{hierarchicalmulticast}. Each \lxb has one or more multicast core nodes associated with it.
A multicast join message at the edge of the RSDN network is directed to the controller for the \lxb of which the receiving switch is a part.  This controller establishes CBT-like state for all members (including the newly joining one) in this \lxb, using one of the \lxb's core nodes.  If the \lxb had previously not had any members in this group, the join is forwarded to the parent \lxb's controller.  The parent controller will then use one if its multicast cores to establish CBT-like state between the newly-joining child and any other children that are part of the group; if the parent has no other children in the group (the notifying one is the first and only), it again forwards the join to \emph{its own} parent.  This proceeds recursively until reaching an \lxb that already has members of the group (and therefore had already informed its parent) or reaching the root \lxb (at which point there is no one else to notify).

Packets addressed to a multicast group are sent to the core within the enclosing lowest-level \lxb, and are broadcast along the tree set up within that \lxb by the CBT-like joins. If the core has been notified by its parent about other members outside of that \lxb, it forwards to its parent core. This recurses, with each core forwarding packets only if it has state indicating that additional members exist in either its parent \lxb or its child \lxbs. Because of RSDN's hierarchical structure, there is always a nearby core for a group, so packets are never sent beyond the highest-level \lxb that encloses all the members of the group. While, for brevity, we do not present evaluation here, the ultimate result is that when group members exhibit geographic locality, there can be a savings in state and latency, particularly when compared to a single core with suboptimal placement.

\paragraph{Anycast:} If one ignores address aggregation, then one can implement anycast merely by announcing the same address from multiple locations and have the unicast route computation find the path to the nearest anycast destination. If one wants to limit the impact on routing state, one can modify the multicast solution described above to provide core-based routing to the nearest anycast destination. In such an approach, each core forwards a packet to only the closest group member rather than to all members.  These distances can be collected via the joins (so that at every point in the hierarchy the core knows the distance to the nearest anycast member via its parent and via each child).

\subsection{Reliability, bootstrapping, security}

All SDN designs must cope with these three issues, and RSDN has no special advantage or disadvantage over other SDN systems in these respects. In terms of reliability in the face of controller failures, we assume that each logical controller is, in reality, a set of controllers that (i) is sufficiently distributed so that they all fail only if most of the components in the associated \lxb have failed, and (ii) uses a coordination system that allows the logical controller to continue operating as long as most (or any) of the constituent controllers remains operational and connected.

RSDN can operate with either an in-band or out-of-band control network, but the in-band case requires some care to successfully bootstrap. RSDN could use a legacy routing algorithm to connect the set of controllers that comprise a logical controller, and the logical controllers to each other.  Integrating this with RSDN need not be difficult (\eg by using a separate set of routing tables), and is a far easier task than general Internet routing as \emph{only the RSDN controllers} need to be routed between using this system.

As for security, we often hear the lament that SDN introduces new threat vectors because an attacker could potentially connect to a switch and gain control over it by pretending to be a controller (or influence a controller by pretending to be a switch).  However, this is not a fundamentally new attack vector: almost all switches have some sort of remote administration capability which, in the hands of an attacker, could be a significant tool to disruption.   Whether SDN or legacy, input to network elements should be authenticated.


\section{Summary}
\label{sec:conclusion}
RSDN is, to our knowledge, the first design that combines the programmability of SDN networks with the hierarchy of legacy networks to create a {\em recursive} framework for global carrier networks. The designs presented in Sections \ref{sec:unicast}, \ref{sec:te}, and \ref{sec:misc} demonstrate that RSDN's programmatic API -- built around two basic event handlers -- provides a clean way to implement a range of unicast, anycast, multicast, and traffic engineering solutions over hierarchical network infrastructures. Moreover, RSDN incorporates a novel network repair mechanism that provides rapid recovery across all such routing functions. We evaluated these designs via simulation on synthetic network topologies that mimic the essential properties of global carrier networks (and we believe these topologies are far superior to the various measured topologies in this respect). While we only presented results from a few of these topologies, we ran simulations over many other topologies and found similar results.

The routing designs presented here are not fundamentally novel in themselves, but they do illustrate that RSDN enables one to achieve good performance (balancing optimality, state, and computation time) and high availability (essentially matching that of the underlying physical connectivity, independent of the particular routing algorithm). Moreover, the approach advocated here could easily be integrated into current SDN controllers (and we have done so with POX~\cite{pox}).

Conceptually, the approach we advocate is complementary to the horizontal scaling we see in existing production SDN controllers: horizontal scale-out provides reliability for a single logical controller, while RSDN's recursive structure leverages network localities to enable algorithms that scale globally but recover locally. Thus, we believe RSDN is a promising approach that complements the current industry directions (of controller scale out) and will enable SDN to control the basic routing functions -- the fabric -- of global carrier networks.

\bibliographystyle{abbrv}
\bibliography{main}

\end{document}